\documentclass[showkeys,twocolumn,showpacs,preprintnumbers]{revtex4}
\usepackage{graphicx}
\usepackage{dcolumn}
\usepackage{textcomp,amsfonts,amsmath,amssymb}%
\usepackage{bm}
\begin{document}

\title{DYNAMIC ORIGIN OF SPECIES}

\author{Michael G.\,Sadovsky}
\affiliation{Institute of computational modelling of SD of RAS;\\ 660036 Russia, Krasnoyarsk.}
\email{msad@icm.krasn.ru}

\altaffiliation{\emph{also at}: Siberian Federal university; 660041 Russia, Krasnoyarsk; Svobodny prosp.,
79.}

\date{\today}
\begin{abstract}
A simple model of species origin resulted from dynamic features of a population, solely, is developed. The
model is based on the evolution optimality in space distribution, and the selection is gone over the
mobility. Some biological issues are discussed.
\end{abstract}



\maketitle

\section{\label{sec:level1}Introduction}
Modelling of the dynamics of biological communities is essential for further understanding both mathematics
\cite{r1,r2,r3,r4}, and biology \cite{z1,z2,z3,z4}. A lot have been done in mathematical ecology and
mathematical population biology since the pioneering works by V.\,Volterra \cite{z5} and A.\,Lotka \cite{z6}.
Yet, there is no comprehensive, steady and solid general theory of the dynamics of biological communities. In
spite of the implementation of the most general and outstanding theorems of the natural selection
\cite{s1,s2,s3}, quite a number of problems still await a researcher to deal with.

Modelling of evolution processes seems to be the only way to figure out various important and significant
details in evolution theory. Since Darwin's famous work \cite{g1}, evolution theory evolved heavily, itself.
A lot have been done in order to introduce the mathematically based methodology into the selection theory;
Haldane'outstanding papers made the most successful start-up, in this direction \cite{haldane1,haldane2}.

Here we propose a simple model of a species origin resulted from the dynamics of a population, solely. An
origin here means a dissociation of an originally uniform population into two subpopulation distinctively
differing in the mobility. This discretion is understood as an appearance of a polymodal distribution of the
beings over the mobility character. To begin with, we should consider, in detail, the model of optimal
migration.

\section{Basic model of optimal migration}\label{bm}
Modelling of spatially distributed communities and populations is quite a problem. Basically, the approach
based on {\sl reaction~\textdiv~diffusion} methodology is used, for modelling of those communities. The
diffusion approach, being quite attractive from mathematical and physical point of view, has nothing to do
with any real system (or even any system just pretending to be relevant to a real one). That is the
diffusion, that makes the problem here.

Diffusion approach gets very strict and absolutely unfeasible constraints on individuals under consideration:
they must transfer in space in random and aimless manner. There is no one species going this way; even
bacteria in continuous cultivation systems control, to some extent, their location. The hypothesis underlying
the diffusion methodology could be improved in neither way.
\begin{figure}
\includegraphics[width=8.8cm]{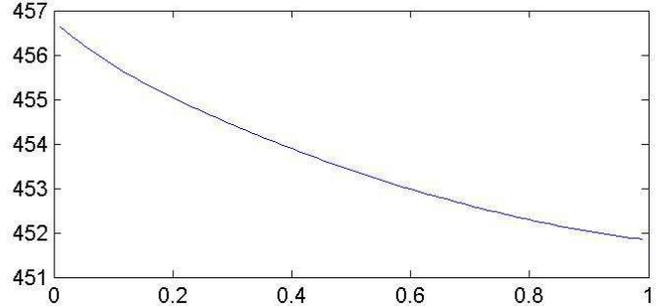}
\caption{\label{fig:1} A distribution resulted from the selective pressure. Here is the distribution in a
single site shown; there are no beings in other one.}
\end{figure}

Thus, the approach based on the principles of evolutionary optimality \cite{s1,s2,s3} opposes the diffusion
methodology; originally, these principles have been introduced by J.B.S.\,Haldane. From the point of view of
the dynamics of spatially distributed communities, these principles force the individuals to migrate in the
manner to increase the \textbf{net reproduction} (NRC). NRC is an average number of descendants (per capita)
determined over a sufficiently long generations chain. In brief, a model based on the the principles should
look like the following. Consider a population occupying two sites; a transfer from site to site makes a
migration. No other movements, inevitable in any real situation, would be taken into account; moreover, we
shall suppose that such movements have no impact on the dynamics.
\begin{figure}
\includegraphics[width=8.8cm]{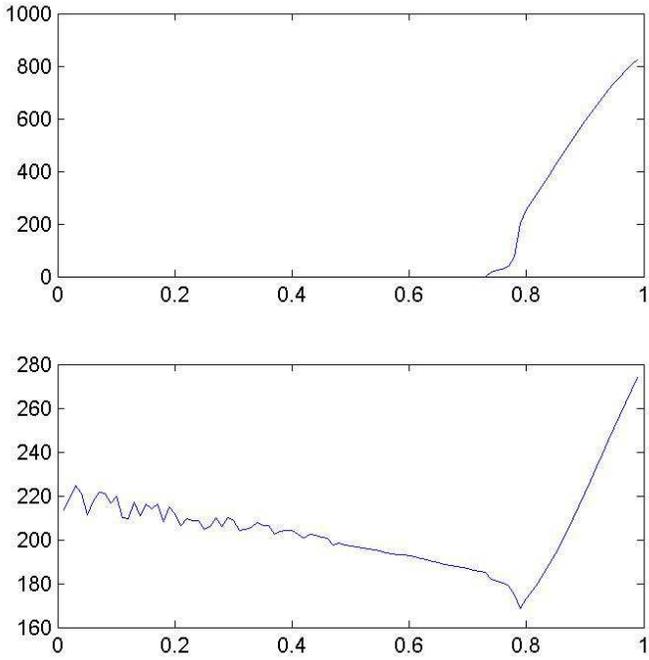}
\caption{\label{fig:3} The case of suppressive selection: the beings of the peculiar mobility class ($p
=0.8$) are eliminated.}
\end{figure}

Suppose, then, that the dynamics of each subpopulation follows the Verchulst's equation:
\begin{subequations}\label{eq1}
\begin{equation}\label{eq1:1}
N_{t+1} = N_t \cdot \left(a - b N_t\right),
\end{equation}
\textrm{and}
\begin{equation}\label{eq1:2}
M_{t+1} = M_t \cdot \left(c - d M_t\right)\,.
\end{equation}
\end{subequations}
Here $N_t$ ($M_t$, respectively) is the subpopulation abundance determined at the time moment $t$. We shall
consider the dynamics in discrete time; continuous time consideration is possible, as well, while it bring
nothing new, but the severe technical difficulties.

We have chosen the Verchulst's equation due to its universality \cite{s4}. The linear functions in
(\ref{eq1}) are NRC, in relevant sites. Thus, a migration from one site to another starts up, as soon, as the
living conditions (measured as a part of NRC) becomes worse {\sl here}, in comparison to similar ones {\sl
there}, with respect to the transfer cost $p$, $0 \leq p \leq 1$. This latter may be considered as a
probability of a successful transfer from a station to another one, with no damage of a further reproduction.
Evidently, an additive pattern of a transfer cost is the simplest one: \[p =
p_{\textrm{out}}+p_{\textrm{in}}+d\,,\] where $p_{\textrm{out}}$ is the cost of a successful leaving of the
site; $p_{\textrm{in}}$ is the cost of a successful intrusion into another site, and $d$ is the pure transfer
cost.

We shall suppose that the individuals under consideration are globally informed; in such capacity, it means
here, that all the parameters (including the transfer cost), as well, as the abundances in each site are
known to them.

The \fbox{N}~$\mapsto$~\fbox{M} migration takes place, as
\begin{equation}\label{eq2}
a-b\cdot N < p \cdot \left(c - d\cdot M\right)\,,
\end{equation}
and the reverse migration takes place, when
\begin{equation}\label{eq3}
c-d\cdot M < p \cdot \left(a - b\cdot N\right)\,.
\end{equation}
The number of migrants must equalize the inequality (\ref{eq2}) (or the inequality (\ref{eq3}),
respectively):
\begin{equation}\label{eq4}
a-b\cdot (N -\Delta) = p \cdot \left(c - d\cdot (M+p\cdot \Delta)\right)
\end{equation}
or
\begin{equation}\label{eq5}
c-d\cdot (M-\Delta) = p \cdot \left(a - b\cdot (N+p\cdot \Delta)\right)\,,
\end{equation}
respectively. Since the number of emigrants may not exceed the total abundance observed within a station,
then, finally, the migration flux $\Delta$ is determined according to
\begin{subequations}\label{eq6}
\begin{equation}\label{eq6:1}
\Delta = \min\left\{N_t, \ \frac{pc-a+dN_t - pdM_t}{b+p^2d}\right\}\,\,\,\phantom{\textrm{.}}
\end{equation}
\textrm{or}
\begin{equation}\label{eq6:2}
\Delta = \min\left\{M_t, \ \frac{pa-c+dM_t - pbN_t}{d+p^2b}\right\}\,.
\end{equation}
\end{subequations}
Here (\ref{eq6:1}) corresponds to \fbox{N}~$\mapsto$~\fbox{M} migration, and vice versa.
\begin{figure}
\includegraphics[width=8.8cm]{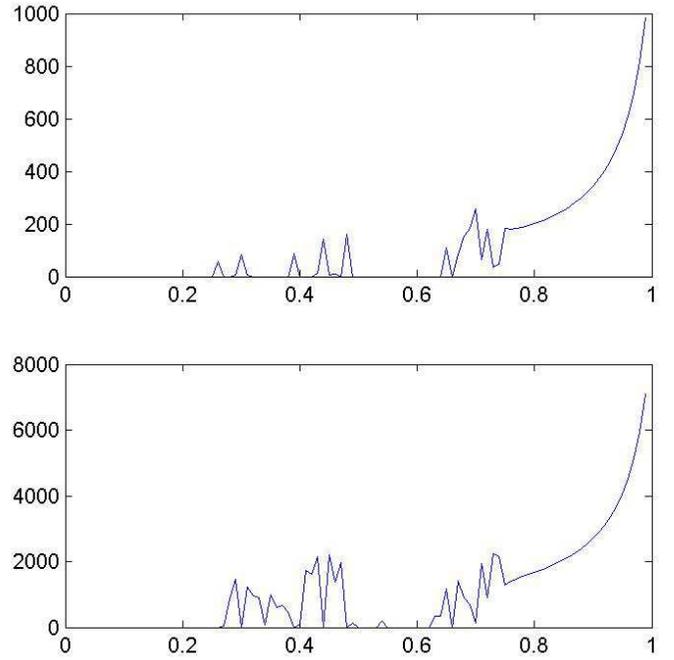}
\caption{\label{fig:2} An occurrence of several subspecies differing in mobility. It is clearly evident, that
some classes of $p$ are eliminated. Upper figure corresponds to \fbox{$N$} site, and lower one to \fbox{$M$}
site.}
\end{figure}

\section{\label{sec:level3}Modification of the basic model}
Suppose, a population (in both sites) is divided into $K$ classes, with respect to the mobility character.
This latter is measured as transfer cost peculiar for each class: \[0 \leq p_1 < p_2 < \ldots < p_{K-1} < p_K
\leq 1\,.\] Let, then, $N_t^{(j)}$ ($M_t^{(j)}$, respectively) be the abundance of of a subpopulation at
$j\textrm{-th}$ class. The dynamics of a subclass (at a subpopulation) is determined according to the
equation:
\begin{subequations}\label{eq7}
\begin{equation}\label{eq7:1}
N_{t+1}^{(j)} = N_{j}^{(j)} \cdot \left\{a - b\cdot \dfrac{1}{K}\displaystyle\sum_{j=1}^K N_{t}^{(j)}
\right\}\,{;}
\end{equation}
\begin{equation}\label{eq7:2}
M_{t+1}^{(j)} = M_{t}^{(j)} \cdot \left\{c - d\cdot \dfrac{1}{K}\displaystyle\sum_{j=1}^K M_{t}^{(j)}
\right\}\,.
\end{equation}
\end{subequations}

Migration rule keeps the same, as (\ref{eq2}) and (\ref{eq3}), for each class, with $p_j$ peculiar for the
class. So, the migration fluxes are determined independently, at each mobility class (i.\,e., for each
$p_j$), the beings relocate themselves between the sites, and finally the reproduction takes place.

\section{Results and Discussion}\label{dickus}
We simulated the dynamics of the system (\ref{eq1}~-- \ref{eq7}), for various parameters. Obviously, the
system is symmetric: nothing will change, if one simulate the dynamics originally for $a=\alpha$, $b=\beta$,
$c=\gamma$ and $d=\theta$, and then do it for $a=\gamma$, $b=\theta$, $c=\alpha$ and $d=\beta$. This symmetry
allows to decrease the dimension of the system. Three parameters control the system: these are $a$, $c$ and
$\lambda = b/d$.

We have simulated the dynamics of the system (\ref{eq1}~-- \ref{eq7}) for $\lambda = 1$, $\lambda = 1/2$,
$\lambda = 1/4$ and $\lambda = 1/10$. The simulation was run in the manner ro scan the parameters $a$ and $c$
area searching the stable polymodal distribution of the the beings over the transfer cost values. Moreover,
in case of $\lambda = 1$, there is a symmetry against the permutation of the parameters of $a$ and $c$.

Further, we show several figures illustrating the observed distributions. Fig.\,\ref{fig:1} shows the
\begin{figure}
\includegraphics[width=8.8cm]{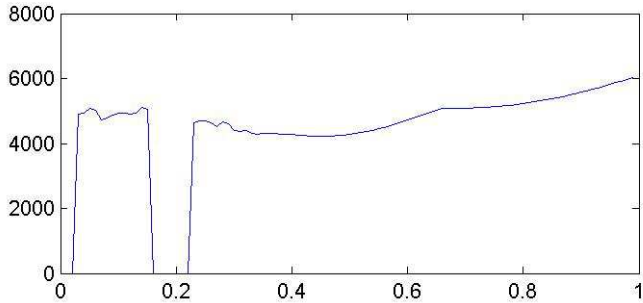}
\caption{\label{fig:4} Another example of disruptive selection.}
\end{figure}
distribution observed at $\lambda = 1/4$ and $a=4$, $c=0.1$. The observed distribution is a typical pattern,
for the case of so called selective pressure: the abundance of the classes with decreasing mobility grows up
monotonously. The figure shows quite rare situation, when the beings with low mobility (small $p$) take an
advantage under the selection pressure.

Figure~\ref{fig:3} shows an interesting case of the suppressive selection: a peculiar mobility class
($p=0.8$). This kind of the distribution might be considered as a partial case of disruptive selection.
Similar situation is shown in Fig.\,\ref{fig:4}; again only one site is shown, since the other one is not
inhabited. This figure is obtained for $\lambda = 1/2$, and $a=0.1$, $c=8.7$.

Figure~\ref{fig:2} shows the situation that is clearly understood as the dissolution of originally
homogeneous (and continuously distributed over the mobility factor $p$) population into two discrete
subspecies. Some classes are completely eliminated due to the selection. Here only few classes have survived;
the survival of the most mobile classes seems rather natural, while the pattern shown in Fig.~\ref{fig:1}
puts on a hypothesis that there are some parameters vales that would yield the survival of the slower beings,
while the faster ones would be eliminated. Finally, Figure~\ref{fig:4} shows the situation of the elimination
of the peculiar classes, while the majority of them survive.

The paper aims to illustrate the occurrence of the selection based on the mobility of organisms, in case of a
non-random migration. Actually, such selection is not a trick: according to the fundamental theorems on the
selection, everything that is inherited, must be processed through the selection \cite{s1,s2,s3,s4}. This
fact explains the results observed at the simulation experiments.

Obviously, one can argue, that there are no species that are present by the beings with all possible
diversity of a parameter (the mobility $p$, in our case). That is right; here we just went the way very
peculiar for a modelling in biology. We have changed the potential diversity of the beings with mobility $p$
that may appear within a population due to mutation for the actual diversity, as if all the mutations already
have been done.

\end{document}